\documentclass{raa}            

\usepackage{graphicx,times}             
\usepackage{natbib}
\usepackage{amssymb,amsmath}
\usepackage{booktabs} 
\usepackage{threeparttable}
\DeclareCaptionLabelSeparator{thinspace}{\kern0.5em} 
\bibpunct{(}{)}{;}{a}{}{,}

\usepackage[pagebackref=true]{hyperref}

\begin{document}

  \title{The Mini-SiTian Array: Light Curves Analysis of Asteroids}

   \volnopage{Vol.0 (20xx) No.0, 000--000}      
   \setcounter{page}{1}          

   \author{Zhaoxing Liu 
     \inst{1,2}
   \and Jian Gao
     \inst{1,2}
   \and Hongrui Gu
     \inst{3}
   \and Yang Huang
     \inst{3,4}
   \and Shaoming Hu
     \inst{5}
   \and Hu Zou
     \inst{3,4}
   \and Keyu Xing
     \inst{1,2}
   \and Hao Huang
     \inst{1,2}
   \and Zehao Zhang
     \inst{1,2}
   }

    \institute{Institute for Frontiers in Astronomy and Astrophysics, Beijing Normal University, Beijing, 102206, China {\it jiangao@bnu.edu.cn}\\
        \and
             School of Physics and Astronomy, Beijing Normal University, Beijing,               100875, People’s Republic of China\\
        \and
             Key Lab of Optical Astronomy, National Astronomical Observatories, Chinese Academy of Sciences, Beijing 100101, People’s Republic of China\\
        \and
             School of Astronomy and Space Science, University of Chinese Academy of Sciences, Beijing 100049, People’s Republic of China\\
        \and
             Shandong Provincial Key Laboratory of Optical Astronomy and Solar-Terrestrial Environment, School of Space Science and Technology, Institute of Space Sciences, Shandong University, Weihai 264209, People's Republic of China\\
\vs\no
   {\small Received 2024 November 12; accepted 2025 March 24}}

\abstract{ The SiTian project, with its vast field of view, will become an ideal platform for asteroid scientific research. In this study, we develop a pipeline to analyze the photometry of asteroids and derive their periods from the data collected by the SiTian pathfinder project Mini-SiTian (MST). The pipeline is applied to the MST f02 region, a MST test region with a sky area of $2.29^{\circ} \times 1.53^{\circ}$. Rotation periods of 22 asteroids are derived by the obtained light curves analysis. Among them, there are 8 asteroids available in the Asteroid Lightcurve Photometry Database (ALCDEF), and 6 of them with more photometric points ($>$200) have similar period parameters as the ones in ALCDEF. Additionally, the periods for 14 of these asteroids are newly obtained and are not listed in ALCDEF. This study demonstrates the feasibility of asteroid photometric research by the SiTian project. It shows that future observations from the SiTian project will provide even more photometry of asteroids, significantly increasing the number of available light curves. The potential vast photometric data of asteroids will help us to further understand the physics of asteroids, their material composition, and the formation and evolution of the solar system.
}

   \authorrunning{Liu ET AL.}            
   \titlerunning{The Mini-SiTian Array: Light Curves Analysis of Asteroids}  

   \maketitle

\section{Introduction}
\label{sect:intro}

Asteroids are vital members of the solar system. They are closely related to the key questions such as the origin of the solar system, the formation of planetary systems, and the origin of life on Earth. In recent years, asteroids have become one of the hot research topics in astronomy and space science with the increasing observation data of asteroids, a series of successful deep space exploration missions to asteroids such as Hayabusa2 (\citealt{2019AcAau.156..387T}) and OSIRIS-Rex (\citealt{2021srml.book..163L}), and the ongoing planetary defense programs like DART (\citealt{2023Natur.616..457C}). 

The study of asteroid light curves is an important aspect of researching asteroid physical properties. Most studies of the rotation characteristics of asteroids are based mainly on lightcurve-derived rotation periods and rotation rates
(\citealt{2002aste.book..113P}).
Concurrently, the physical parameters such as shape can be analyzed
(\citealt{2001Icar..153...24K, 2015aste.book..745S, 2015aste.book..183D, 2024MNRAS.528.3523F}).
Moreover, large data volume studies of asteroid light curves can conduct statistical analyses on the distribution of asteroid rotation periods 
(\citealt{2020ApJS..247...26P}),
all main-belt asteroids between 0.4 and 10 km have rotation rates less than 2.2h
(\citealt{2009Icar..202..134W}),
the characteristics of rotation periods in different asteroid families (\citealt{2022A&A...661A..48S}).
Even more, it can provide deeper insights into the evolution of the solar system
(\citealt{2014Natur.505..629D}).
Furthermore, the activity of asteroids can also be studied a lot with its physical parameters
(\citealt{2004AJ....127.2997H, 2019NatSR...9.5492S})
and even exploring the origin of water on Earth
(\citealt{2019Natur.568...55L}).

Now, with the rise of time-domain astronomy, numerous wide-field time-domain sky surveys are being promoted both domestically and internationally, simultaneously presenting opportunities for research on asteroids. Domestic wide-field survey facilities, such as the Tsinghua University-Ma Huateng Telescopes for Survey (TMTS) 
(\citealt{2020PASP..132l5001Z}),
the Multi-channel Photometric Survey Telescope of Yunnan University (Mephisto)
(\citealt{2019gage.confE..14L}),
the “Mozi” Wide Field Survey Telescope (WFST)
(\citealt{2016SPIE10154E..2AL}),
and so on, all possess the potential capability for asteroid photometric study (\citealt{2023MNRAS.521.3925X}). In this work, we focus on conducting asteroid photometry with the SiTian project
(\citealt{2021AnABC..93..628L}),
with plans to extend this effort to other wide-field surveys in the future.

The SiTian project consists of approximately 20 groups, each equipped with three 1-meter telescopes. SiTian will scan at least 10,000 square degrees of sky using three bands in half an hour, down to a 5-$\sigma$ detection limit of $V \approx 21$ mag. This large-field observation mode is highly suitable for conducting research on asteroids in the solar system (\citealt{2021AnABC..93..628L}).

As the pathfinder project of the SiTian project, Mini-SiTian (MST) is equipped with three 30-centimeter catadioptric Schmidt telescopes named Mini-001 (MST1), Mini-002 (MST2), and Mini-003 (MST3). Each telescope is equipped with a commercial-grade Complementary Metal-Oxide-Semiconductor (CMOS) sensor, the SONY IMX455, with $9576 \times 6388$ pixels. Covering a $2.29^{\circ} \times 1.53^{\circ}$ field of view (FOV) and the pixel scale is $\text{0.862}^{''}$/pixel. MST1, MST2, and MST3 are equipped with filters similar to the SDSS system filters $i^{'}$, $g^{'}$, and $r^{'}$, respectively. \textcolor{red}{(Xiao et al. 2025; He et al. 2025)}. In this series of works, \textcolor{red}{Zhang et al. (2025)} demonstrated that the dark current and readout noise of the MST's CMOS camera are negligible, with the dark current approximately 0.002 $\text{e}^{-}$ $\text{pixel}^{-1}$ $\text{s}^{-1}$ at 0°C and the readout noise below 1.6 $e^-$. \textcolor{red}{Xiao et al. (2025)} showed that the MST's zero-point correction accuracy is $\sim$ 1 mmag, with an uncertainty of 4 mmag for G $\sim$ 13 mag. MST's CMOS achieves photometric accuracy comparable to CCDs under current exposure conditions. During the test observation period of MST, three regions were chosen to test the performance of long-duration continuous observations, which were named f01, f02, and f03. Among them, the f02 region is located at $\alpha = 72.947^{\circ}$ and $\delta = +41.015^{\circ}$ and has the richest data from observations \textcolor{red}{(Gu et al. 2025)}. Using the observations of the f02 region, we conduct a study on the light curves of asteroids, in order to demonstrate the feasibility of MST for asteroid light curve research.

The structure is as follows: Sec.~\ref{sect:Data} briefly introduces the data and methods. The photometric results of the asteroids are presented in Sec.~\ref{sect:Result}. A comparison of the results with existing databases is provided in Sec.~\ref{sect:Discussion}, and the summary is given in Sec.~\ref{sect:conclusion}.

\section{Data and Method}
\label{sect:Data}

\subsection{Observation}
\label{sect:observation}

Observations of the f02 region began on December 24, 2022, and ended on February 6, 2023. MST2 with SDSS-g band and MST3 with SDSS-r band were used in this observation. 3636 images were taken by MST2 and 3554 images were taken by MST3 for the f02 region. The exposure time for each observation was 300 seconds \textcolor{red}{(Gu et al. 2025)}. Based on the actual observations, the full width at half maximum (FWHM) for the frames during each night is mostly between $2^{''}$ and $3^{''}$. However, the FWHM of each frame also varies with the seeing conditions at the site, which may change during the night. The worst FWHM is mostly between $3^{''}$ and $4^{''}$. The 3-$\sigma$ magnitude limit for a single image could reach 20.0 mag in the $g^{'}$-band under the best observational conditions, and 19.4 mag in the $r^{'}$-band \textcolor{red}{(He et al. 2025)}.

During continuous observations, the positions of asteroids vary over time, which photometric may be affected by background stars. To ensure the accuracy of subsequent analysis, we use the image differencing method to effectively reduce the influence of background stars. \textcolor{red}{Gu et al. (2025)} developed the Mini-Sitian Real-Time Image Processing system (STRIP) and completed the preprocessing and image subtraction of the images from f02 field. Raw images were processed with bias-field subtraction and flat-field division by STRIP. STRIP used the World Coordinate System (WCS) solving tool from ‘astrometry.net’ to compute the WCS for all images and then used Scamp (\citealt{2006ASPC..351..112B}) to obtain a more precise WCS solution. A WCS template was created by STRIP, and all images were aligned to this template using SWarp (\citealt{2010ascl.soft10068B}). A template image was produced by stacking past high-quality observations. Finally, the subtraction images were produced by differencing them against the template image using hotpants (\citealt{2015ascl.soft04004B}).

\begin{figure}
       \centering
       \includegraphics[width=\textwidth, angle=0]{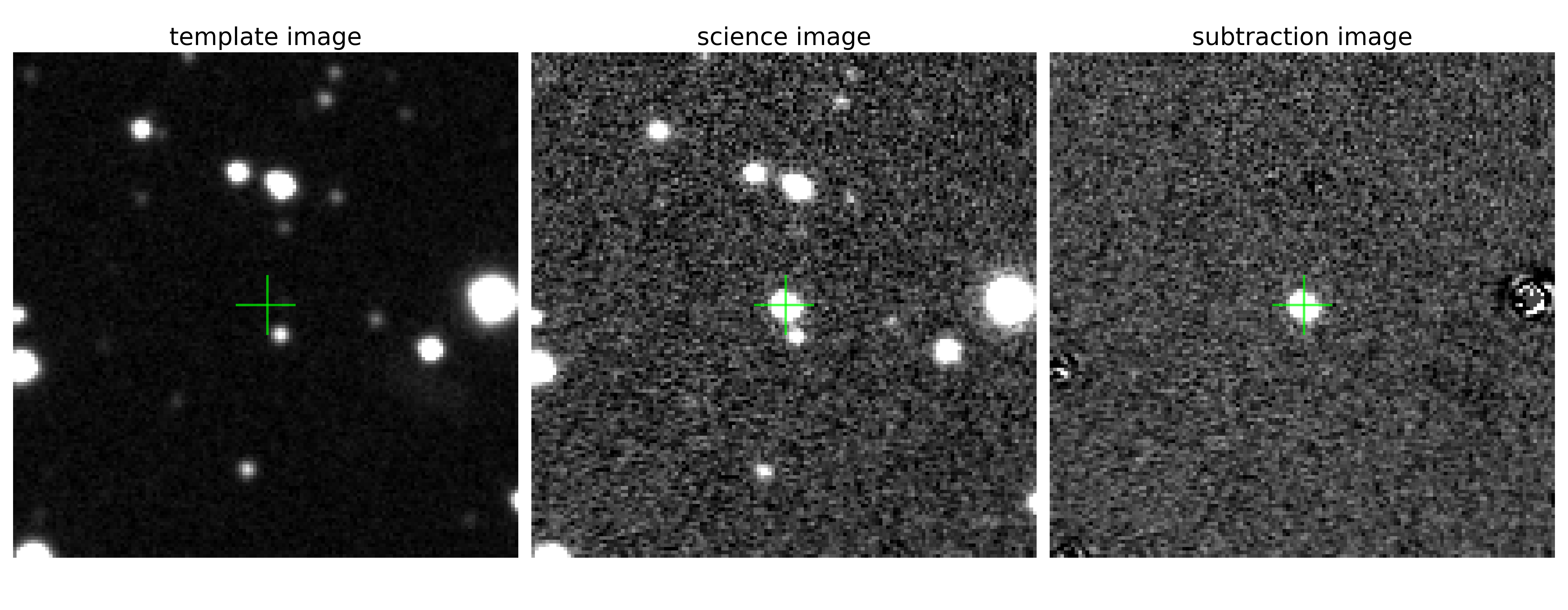}
       \caption{Comparison between the reference image and a single image in the f02 region taken by MST2. Their subtracted image is shown in the right panel. The center of the subtraction image is the asteroid 1524 Joensuu.}
       \label{Fig1}
    \end{figure}

Figure~\ref{Fig1} shows an example of the difference between the template image, science image, and residual image. The asteroid does not appear in the template image because after using 3-$\sigma$ clipping method, moving objects will be removed in the template image. Therefore, asteroid photometry using the subtracted image can eliminate most of the interference from background stars.

In this series of works, \textcolor{red}{Gu et al. (2025)} utilize BP/RP (XP) spectra from Gaia DR3 in conjunction with the filter transmission curves of the MST array to obtain two-band theoretical magnitudes for stars within the MST filter system. After coadding template images of the f02 field, aperture photometry is performed using \textit{SExtractor} (\citealt{1996A&AS..117..393B}), producing a catalog for the f02 field. This catalog is then cross-matched with the catalog derived from XP spectra, generating a table that associates aperture photometry values (measured in Analog-to-Digital Units (ADU)) with theoretical magnitudes from Gaia XP spectra. By taking the logarithm of the ADU values and multiplying by -2.5, ADU values are converted into instrumental magnitudes. These instrumental magnitudes are subtracted from the theoretical magnitudes from the XP spectrum to determine the zero points of the template images. A 3-$\sigma$ clipping is applied to the results, and the methodologies are separately applied to template images from both MST2 and MST3, each with a distinct filter, yielding two unique zero points: 28.03 mag (MST2) and 27.30 mag (MST3).

\subsection{The Asteroid Ephemeris}
\label{sect:ephemeris}

To perform photometry on asteroids, it is essential to identify them in the images and determine their positions. For this purpose, we download the catalog of all known asteroids in the Solar System from the Lowell website \footnote{\textless https://asteroid.lowell.edu\textgreater Asteroid Orbital Elements Database. 2024.}
(\citealt{2022A&C....4100661M}),
which contains high-precision orbital elements of asteroids and other related parameters, such as the absolute magnitude \textit{H}
(\citealt{1989aste.conf..524B}). We use the data version of January 5, 2024, which includes information on 1,340,596 asteroids.

Next, we use the \textit{Aleph} software package\footnote{https://github.com/josepenaz/ephs} to compute the $\alpha$ and $\delta$ of asteroids. \textit{Aleph} integrates with the Lowell Asteroid Orbital Elements Database and directly reads the orbital elements of asteroids from the \textit{astorb.dat} file. Through the integral of orbit, it calculates the positions of asteroids in the International Celestial Reference System (ICRS) at different times. We provide four key parameters to \textit{Aleph}: the equatorial coordinate of the FOV center ($\alpha = 72.947^{\circ}$ and $\delta = +41.015^{\circ}$), the search range (a conical search centered on the FOV center with a diameter of $1.5^\circ$),  the Julian Day (JD) at which the observation images were taken, and the station coordinates of the Xinglong Station of the National Astronomical Observatories, where the MST is located. Based on these parameters, \textit{Aleph} calculates the positions of asteroids within the search area at the time of each image captured. Finally, it obtains the precise $\alpha$ and $\delta$ of the target asteroids.

Based on the quality of the images, we select asteroids with $V \leq 20 \, \text{mag}$. Using the WCS \footnote{https://docs.astropy.org/en/stable/wcs/index.html} information of the images, we convert the $\alpha$ and $\delta$ of the target asteroids into pixel coordinates for each image. Finally, we obtain the asteroid ephemeris table (see Table~\ref{Tab1}), which includes asteroid number and name, JD, ICRS coordinate ($\alpha$ and $\delta$), the number of MST, and pixel coordinates. From the MST2 data, we identify 26 asteroids, 24 of which are main-belt asteroids and 2 of which are Trojan objects. From the MST3, we identify 24 asteroids, 22 of which are main-belt asteroids and 2 of which are Trojan objects. The reason why MST3 finds fewer asteroids than MST2 is that fewer images are taken by MST3.

\begin{figure}
       \centering
       \includegraphics[width=\textwidth, angle=0]{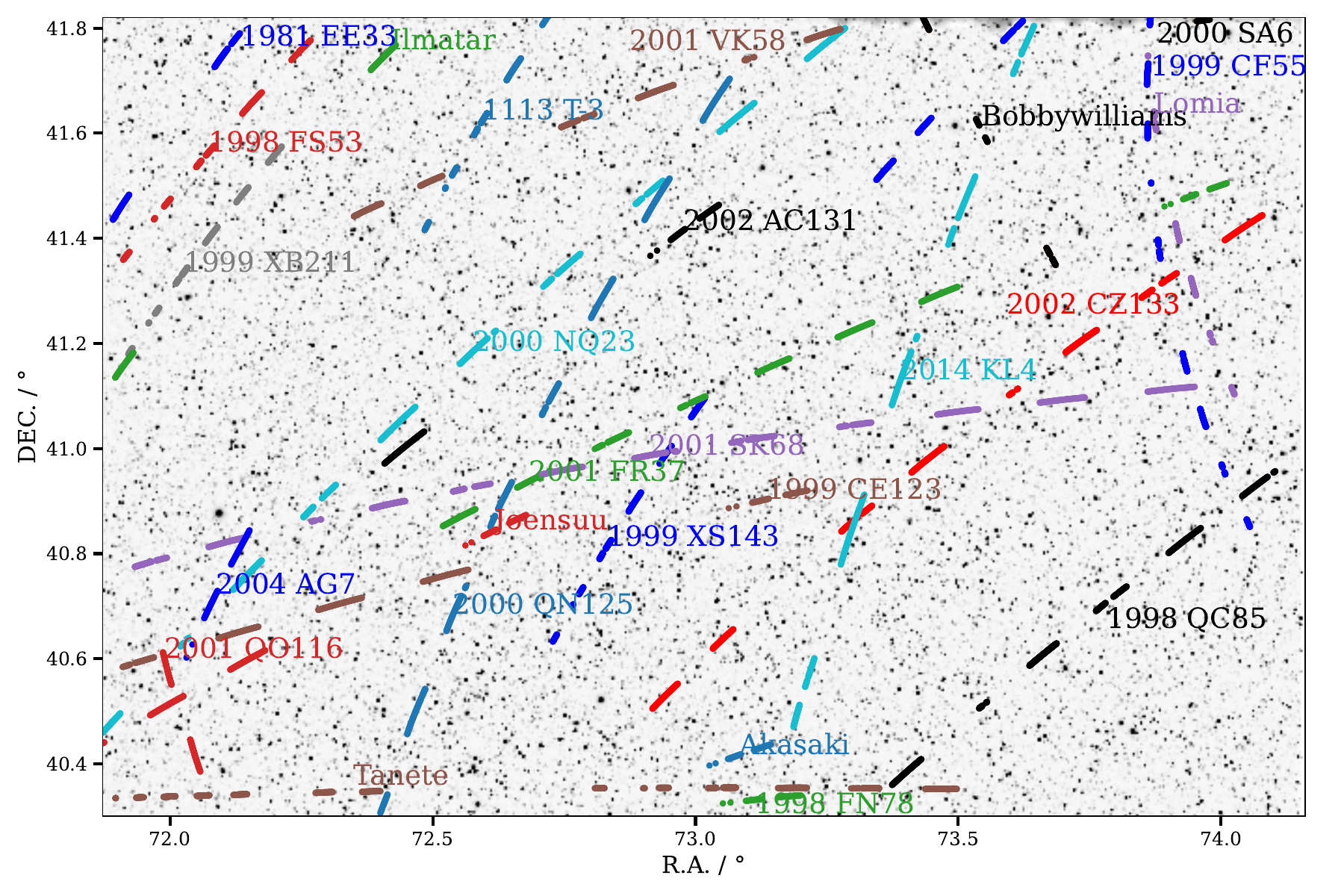}
       \caption{An example image of MST2 with the identified asteroids. The background is a real sky image taken by the MST2, and the colored dots represent the tracks of the asteroids in the asteroid ephemeris, which appear as dashed lines because they can only be observed at night. The colored labels are the names of the asteroids.}
       \label{Fig2}
    \end{figure}

We incorporated the positional coordinates of the asteroids from each image taken by MST2 into a scatter plot, where each asteroid in each image is represented by a point. As the position of the asteroid changes across consecutive images, these points connect to form tracks, with the length of the tracks reflecting the number of images taken. Figure~\ref{Fig2} shows the identified asteroids in an original image taken by MST2, clearly indicating that the asteroids were passing through the FOV during the MST f02 observation.

    \begin{table}[!ht]
        \centering
        \caption{Asteroid Ephemeris in the F02 test region}\label{Tab1}
        \begin{threeparttable}
        \begin{tabular}{*{8}{c}} 
        \toprule
            \textbf{Number} & \textbf{Name} & \textbf{JD[day]} & \textbf{$\alpha$[deg]}\textsuperscript{a} & \textbf{$\delta$[deg]}\textsuperscript{a} & \textbf{MST-number} & \textbf{x-pixel} & \textbf{y-pixel} \\ \midrule
            772 & Tanete & 2459962.965185 & 72.91261 & +40.30397 & 2 & 5033.74 & 228.49 \\ 
            772 & Tanete & 2459962.972558 & 72.91131 & +40.30410 & 2 & 5029.56 & 228.52 \\ 
            772 & Tanete & 2459962.976250 & 72.91065 & +40.30416 & 2 & 5027.46 & 228.54 \\ 
            \multicolumn{8}{c}{\dots} \\ 
            39433 & 1113 T-3 & 2459963.252211 & 72.45967 & +41.60614 & 2 & 2980.04 & 5466.34 \\ 
            39433 & 1113 T-3 & 2459963.255903 & 72.45931 & +41.60577 & 2 & 2979.11 & 5464.69 \\ 
            39433 & 1113 T-3 & 2459963.259595 & 72.45895 & +41.60541 & 2 & 2978.18 & 5463.03 \\ 
        \bottomrule
        \end{tabular}
        \begin{tablenotes} 
            \item[a] The $\alpha$ and $\delta$ are the ICRS coordinates.
        \end{tablenotes}
        \end{threeparttable}
    \end{table}

\subsection{Data Analysis}
\label{sect:analysis}

\begin{figure}
       \centering
       \includegraphics[width=\textwidth, angle=0]{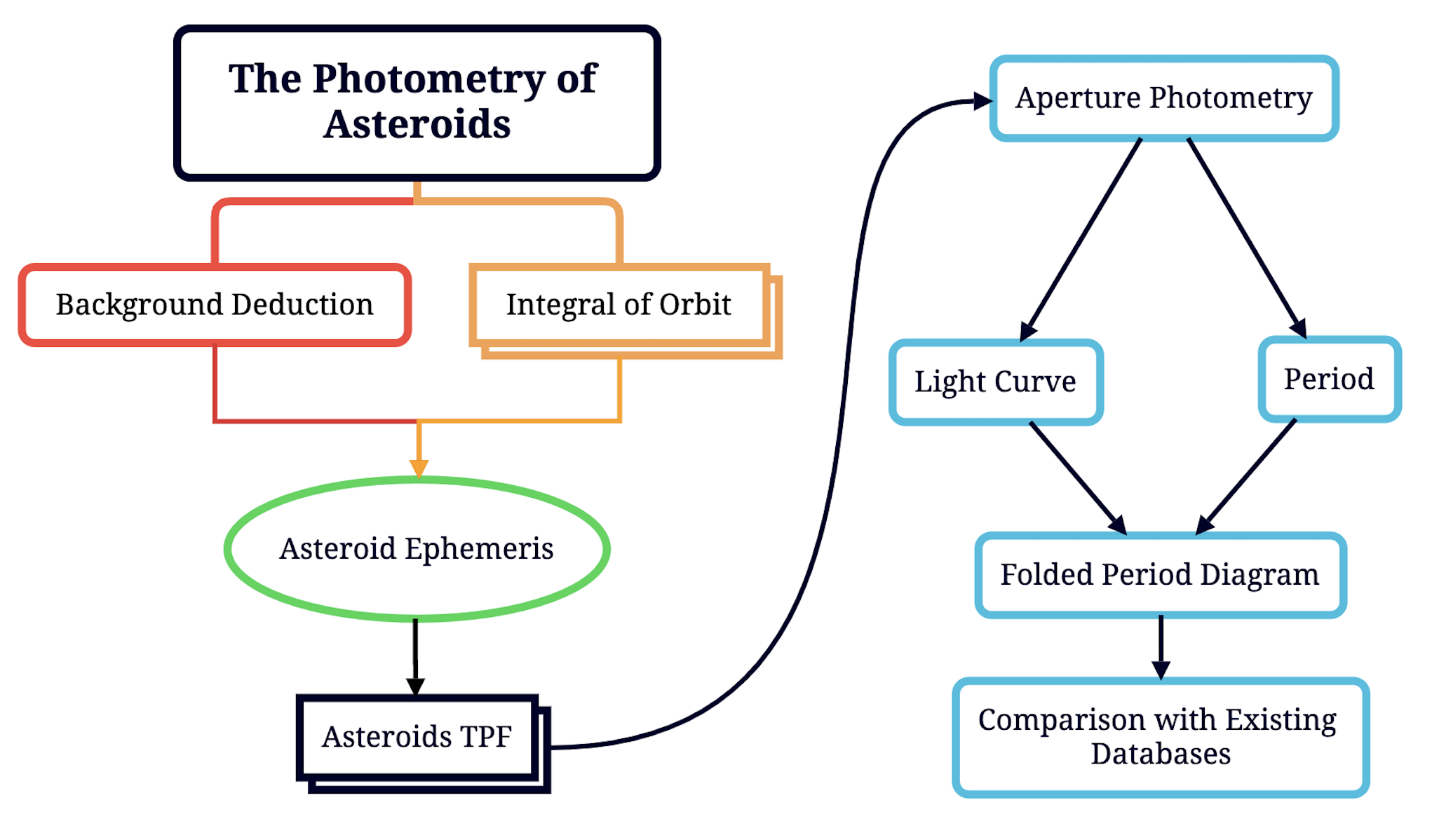}
       \caption{Flowchart of the asteroid photometry for the MST.}
       \label{Fig3}
    \end{figure}

As shown in Figure~\ref{Fig3}, we provide a detailed introduction to the data processing pipeline for the asteroid photometry in the MST project.

\subsubsection{Image Cropping}

Based on the asteroid ephemeris table, we extract the observation times for each asteroid and match them with the images to precisely locate the pixel coordinates of the target asteroids. Using the \textit{Cutout2D} package in \textit{Astropy}\footnote{https://github.com/spacetelescope/astrocut}
(\citealt{2022ApJ...935..167A}),
we crop the subtraction image to $100 \times 100$ pixels
 Target Pixel Files (TPFs) centered on each asteroid.

After cropping, we obtain 13,218 TPFs for the f02 region from MST2 and 9,447 TPFs from MST3. Through image quality analysis, we find that some images are unsuitable for photometric measurement. Due to the instability inherent in image subtraction methods, the presence of saturated pixels, and variable stars appearing in the subtraction images, asteroids may be obscured. Some asteroids are not fully captured because they are on the edge of the image, and some low-quality images make it difficult to identify asteroids. During photometry, we further eliminate affected images to ensure the accuracy of the analysis. This includes images where asteroids cannot be found in the TPFs, images with significant photometric deviations (more than 3-$\sigma$), or images with other issues.

\subsubsection{Aperture Photometry}

After obtaining the TPFs, we find that some targets are not precisely centered in the images because the orbit integration is not accurate enough. To achieve more accurate photometry, we use the \textit{DAOStarFinder} program
(\citealt{1987PASP...99..191S}) from the \textit{photutils} package\footnote{https://photutils.readthedocs.io/en/stable/api/photutils.detection.DAOStarFinder.html}(\citealt{2023zndo...1035865B}) to realize star finding in the TPFs for precise localization of the asteroids. We conduct aperture photometry using the localization center as the center of the aperture. Based on the MST observation data described in Sec~\ref{sect:observation}, we set the FWHM to $3.448^{''}$ (4 pixels) and use 1.5 times the FWHM as the radius for the \textit{CircularAperture} and \textit{aperture photometry} methods\footnote{https://photutils.readthedocs.io/en/stable/aperture.html} from the \textit{photutils} package. While obtaining the total flux of the asteroids, we construct two large circles with radii of 5 times and 7 times the FWHM, centered on the localization center, and detect the total flux values within these large circles. The average flux value of all pixels in the annulus between the two outer circles is calculated and termed as the sky background flux (\textit{Msky}). The target flux is obtained using the formula $\textit{Flux} = \textit{Sum} - \text{Area} \times {\textit{Msky}}$.

The relationship between ADU values and magnitudes, as derived from the established zero points (See Sec~\ref{sect:observation}), is shown in Equation~\eqref{eq1}.

    \begin{subequations}\label{eq1}
      \begin{align}
        \textit{Mag} = 28.03 - 2.5 \times \lg(\textit{flux}) \quad \quad \quad \text{ (MST2)} \label{eq:Mst2} \\
        \textit{Mag} = 27.30 - 2.5 \times \lg(\textit{flux}) \quad \quad \quad \text{ (MST3)} \label{eq:Mst3}
      \end{align}
    \end{subequations}

Based on the readout noise, dark current, and gain values provided by \textcolor{red}{Zhang et al. (2025)}, this study estimates asteroid photometric errors by considering Poisson noise, skylight background noise, readout noise, and dark current noise. For bright targets (G $\sim$ 12 mag), the uncertainty is around 2 mmag, but it increases to 0.6 mag for fainter targets at the dim end. For most targets around magnitude 18, the uncertainty is approximately 0.1 mag.

Due to the finite speed of light and the asteroid's motion relative to the observation station, aberration must be taken into account. The heliocentric Cartesian coordinates (\textit{x, y, z}) of the asteroid at each epoch can be calculated using the \textit{Aleph} software package (see Sec~\ref{sect:ephemeris}). By combining these coordinates with the station's location (obtained from \textit{astropy.coordinates}), an iterative algorithm is employed to correct the impact of light aberration on the observed epoch.

To improve period analysis, the observed magnitude of the asteroid was converted to absolute magnitude (\textit{H}) using the formula from \citet{1989aste.conf..524B} (see their equation (36)), considering distance and the phase angle $(\alpha)$ corrections.

\section{Result}
\label{sect:Result}

\begin{figure}
       \centering
       \includegraphics[width=\textwidth, angle=0]{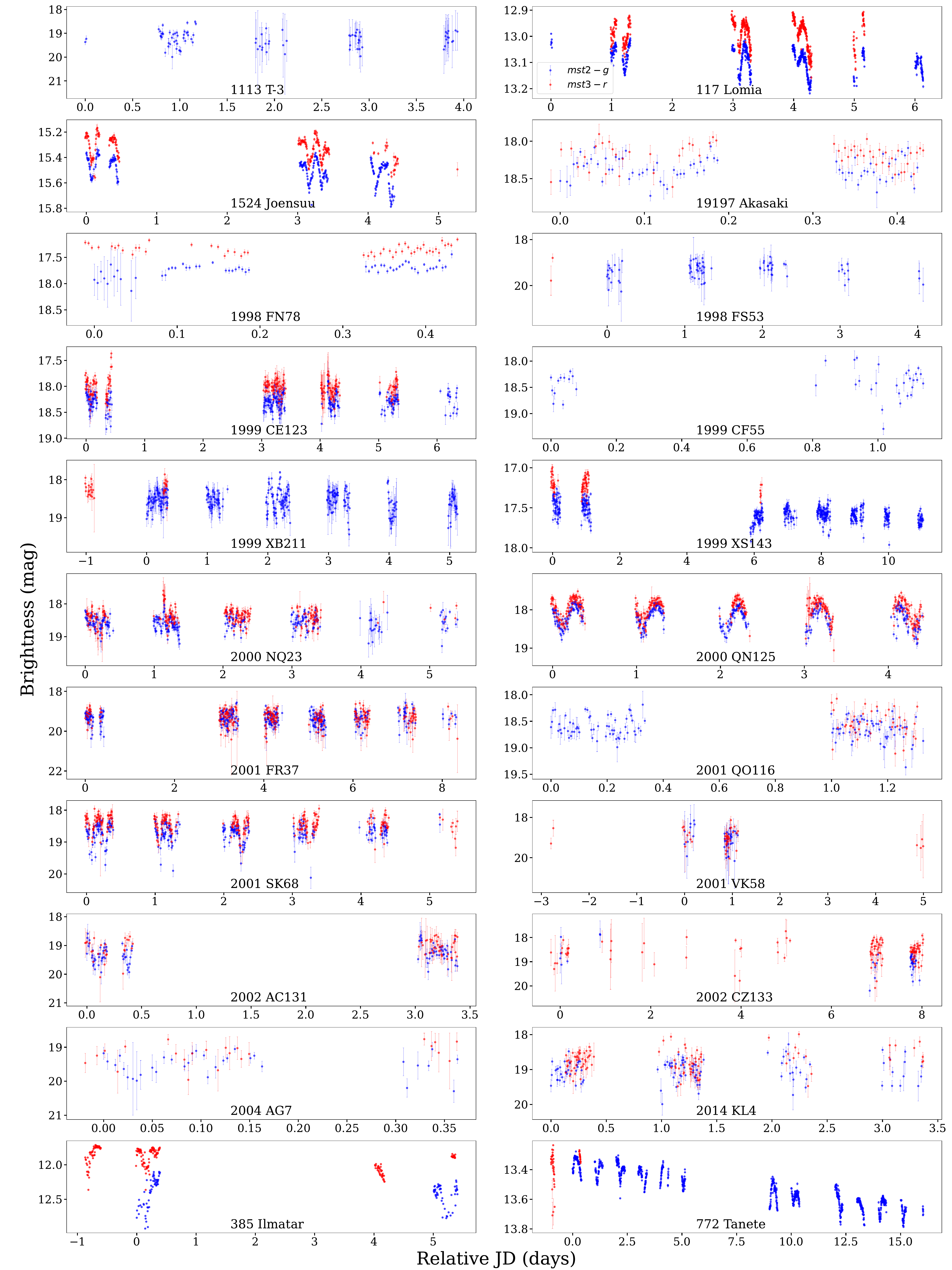}
       \caption{Light curves of 22 asteroids. Blue points are photometric points obtained using MST2 observations with \textit{g}-band, and red points are obtained using MST3 with \textit{r}-band.}
       \label{Fig4}
    \end{figure}

Figure~\ref{Fig4} presents the light curves of 22 asteroids obtained through accurate aperture photometry in the g-band and r-band (with r-band data unavailable for two asteroids), using epoch data corrected for aberration and original observed magnitudes.

\begin{figure}
       \centering
       \includegraphics[width=\textwidth, angle=0]{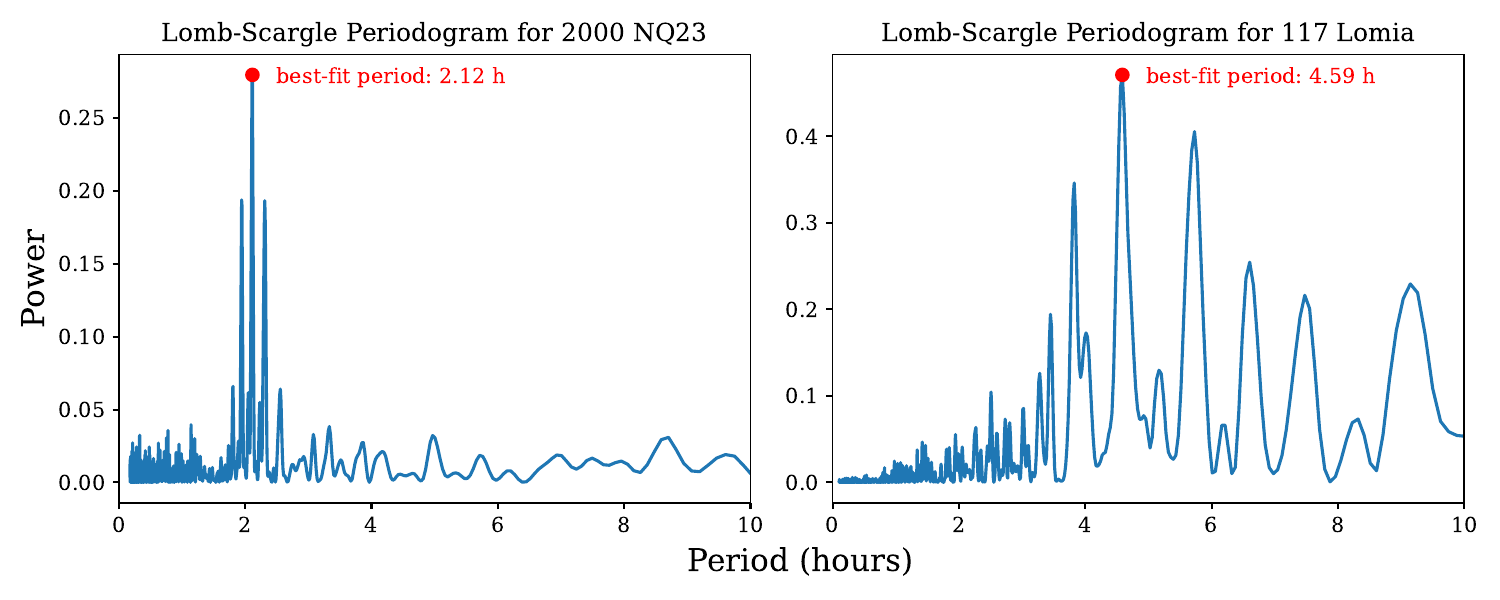}
       \caption{Two examples of the periodogram by the LS. According to this figure, the best-fit LS period of 2000 NQ23 is 2.12h (double best-fit LS period is 4.24h), and the best-fit LS period of 117 Lomia is 4.59h (double best-fit LS period is 9.18h).}
       \label{Fig5}
    \end{figure}

Due to the non-uniform sampling of photometry points, we use the Lomb-Scargle (LS) analysis technique\footnote{https://docs.astropy.org/en/stable/timeseries/lombscargle.html} 
(\citealt{1976Ap&SS..39..447L, 1982ApJ...263..835S}) to derive the best-fit period. In the LS implementation, we followed 
\citet{2019ApJS..245...29M} and \citet{2023PSJ.....4...61L} and set the asteroids period at twice the best-fit LS period. The reason is that we assume the light variation of the asteroid is caused by the rotation. Besides, the asteroid should have a double-peaked light curve generated by the rotation of an elongated body (\citealt{2020ApJS..247...26P}). It shows in Figure~\ref{Fig5} that two examples of the periodogram by the LS.

\begin{figure}
       \centering
       \includegraphics[width=\textwidth, angle=0]{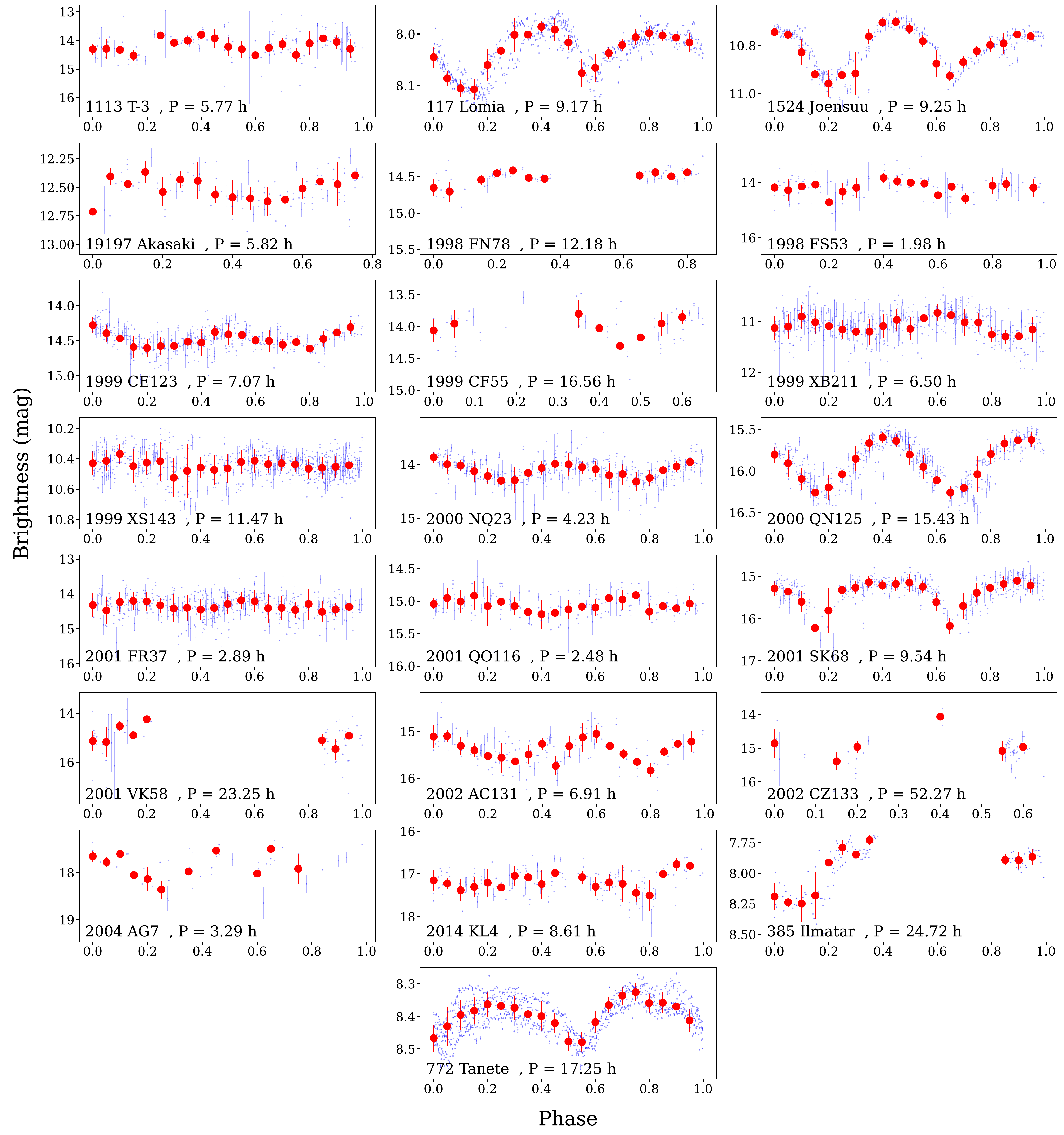}
       \caption{The folded light curves of the 22 asteroids. Brightness is the absolute magnitudes (H) of asteroids, the blue points are the original photometric points, and the red points are the average photometric values in equal intervals.}
       \label{Fig6}
    \end{figure}

It shows in Figure~\ref{Fig6} that the folded light curves with the best-fit rotation period for the 22 asteroids detected by the \textit{g}-band MST2 which have more data than MST3. Using each asteroid's best-fit period as a reference, we fold the light curves and divide one period into 20 equal parts. Here, epoch data corrected for aberration and magnitudes converted to absolute magnitude \textit{(H)} are used. Because some asteroids have too few photometric points to cover the whole period, only the portion covered by the photometric points is counted such as 2001 VK58, 2002 CZ133, and so on. 

\begin{figure}
       \centering
       \includegraphics[width=\textwidth, angle=0]{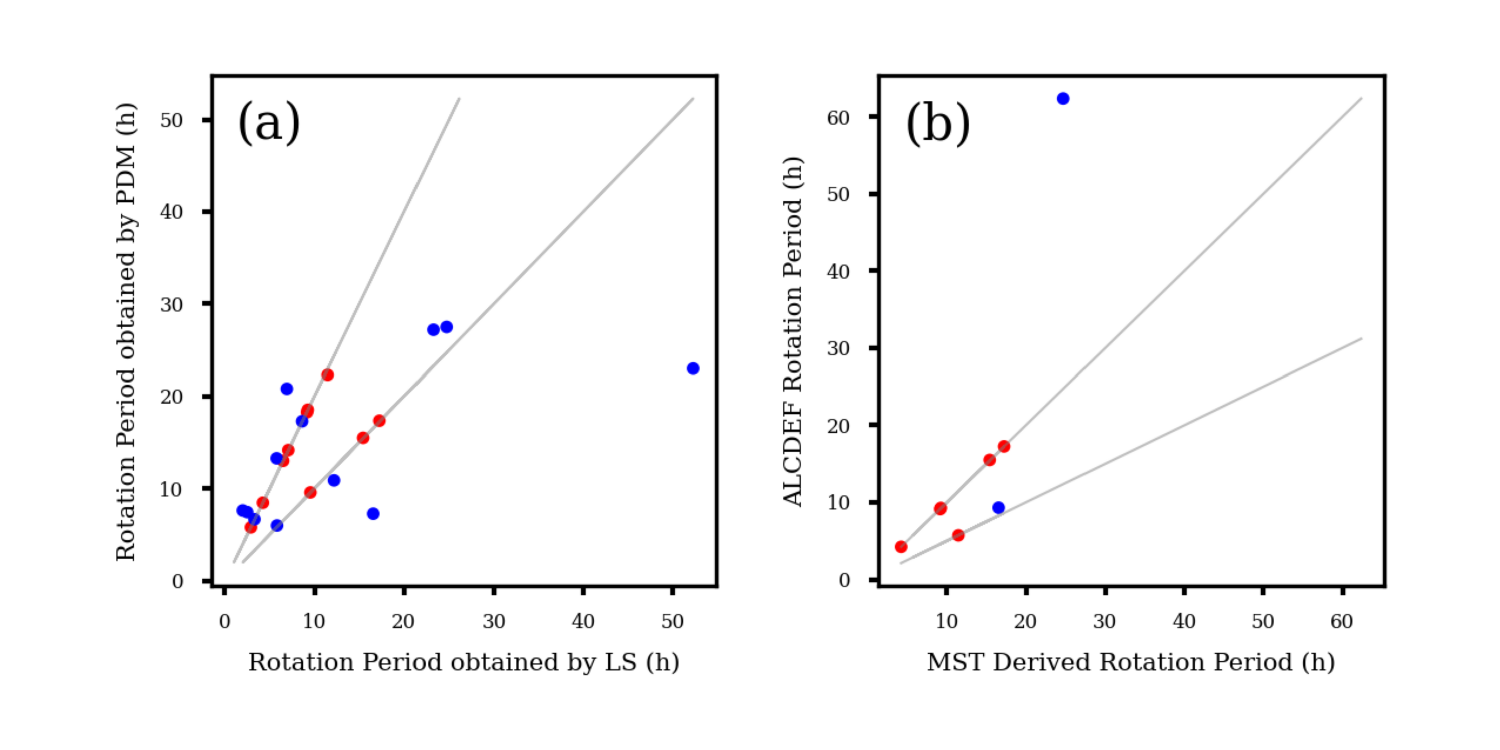}
       \caption{(a) Comparison of the period results for 22 asteroids obtained by LS and PDM. \\
       (b) Comparison of the period results for 22 asteroids obtained by MST and ALCDEF.\\
       The red points are the asteroids with more than 200 photometric points, and the blue points are the asteroids with less than 200 photometric points. The two reference lines of 1:1 and 2:1 are marked with gray lines.}
       \label{Fig7}
    \end{figure}

To ensure the accuracy of periodic analysis, we also apply the “phase dispersion minimization” (PDM)\footnote{https://pyastronomy.readthedocs.io/en/latest/pyTimingDoc/pyPDMDoc/pdm.html} method (\citealt{1978ApJ...224..953S, 2008ssbn.book..129S, 2012RAA....12.1714W}). For better comparison and to emphasize the double-peaked property of asteroid light curves, the period derived from the PDM method is also multiplied by two. It shows in Figure~\ref{Fig7}(a) that the relationship between the period measured by LS and PDM. Since assume that the asteroid light curve has a double-peaked property, the periods of 1:1 or 1:2 by two methods are credible. The period obtained for the asteroids with more than 200 photometric points is similar, and the one obtained with less than 200 light points is not stable due to inadequate data.

In this work, the rotation period uncertainty is estimated using half the height of the main peak in the periodogram (\citealt{2018ApJS..236...16V}). Specifically, in the LS or PDM analysis, the position of the main peak is identified, and the half-width at half-maximum of the main peak is used to quantify the error during the period measurement. The method achieves a maximum precision of 0.01 hours. All the period and error data can be viewed in Table~\ref{Tab2}.

    \begin{table}[!ht]
        \centering
        \caption[]{ The parameters of the identified asteroids in this work }\label{Tab2}
        \begin{threeparttable} 
        \begin{tabular}{*{10}{c}} 
        \toprule
            \textbf{Name} & $\textit{\textbf{V}}\textbf{[mag]}$\textsuperscript{a} & $\textit{\textbf{g}}\textbf{[mag]}$\textsuperscript{b} & $\textit{\textbf{r}}\textbf{[mag]}$\textsuperscript{b} & \textbf{\textit{g - r}}\textsuperscript{b} & 
            $\textbf{\textit{V}}_\textbf{MST}\textbf{[mag]}$\textsuperscript{c}&
            $\textbf{\textit{P}}_\textbf{LS}\textbf{[h]}$\textsuperscript{d}& $\textbf{\textit{P}}_\textbf{PDM}\textbf{[h]}$\textsuperscript{d} & $\textbf{\textit{P}}_\textbf{a}\textbf{[h]}$\textsuperscript{e} & \textbf{Points}\\ \midrule
            772 Tanete & 13.58 & $13.56\pm0.002$ & $13.37\pm0.007$ & 0.19 & 13.43 & $17.25^{+0.14}_{-0.14}$ & $17.35^{+0.15}_{-0.16}$ & 17.258 & 918 \\ 
            117 Lomia & 13.23 & $13.09\pm0.002$ & $12.99\pm0.003$ & 0.1 & 13.01 & $9.17^{+0.17}_{-0.16}$ & $18.27^{+0.72}_{-0.41}$ & 9.127 & 569 \\ 
            1999 XS143 & 17.60 & $17.57\pm0.04$ & $17.21\pm0.05$ & 0.36 & 17.35 & $11.47^{+0.09}_{-0.11}$ & $22.30^{+1.08}_{-1.72}$ & 5.72 & 489 \\ 
            2000 QN125 & 17.78 & $18.19\pm0.06$ & $17.98\pm0.09$ & 0.21 & 18.05 & $15.43^{+0.46}_{-0.44}$ & $15.49^{+0.46}_{-0.39}$ & 15.502 & 354 \\ 
            1999 XB211 & 18.27 & $18.57\pm0.21$ & $18.24\pm0.18$ & 0.33 & 18.37 & $6.49^{+0.11}_{-0.07}$ & $13.01^{+0.16}_{-0.14}$ & - & 345 \\ 
            2001 FR37 & 19.43 & $19.37\pm0.21$ & $19.32\pm0.33$ & 0.05 & 19.32 & $2.89^{+0.19}_{-0.09}$ & $5.78^{+0.01}_{-0.01}$ & - & 321 \\ 
            2000 NQ23 & 18.52 & $18.57\pm0.12$ & $18.39\pm0.16$ & 0.18 & 18.45 & $4.23^{+0.03}_{-0.04}$ & $8.45^{+0.06}_{-0.08}$ & 4.228 & 298 \\ 
            1999 CE123 & 18.13 & $18.30\pm0.09$ & $18.03\pm0.14$ & 0.27 & 18.13 & $7.07^{+0.09}_{-0.07}$ & $14.16^{+0.17}_{-0.12}$ & - & 273 \\ 
            2001 SK68 & 18.61 & $18.71\pm0.11$ & $18.44\pm0.15$ & 0.27 & 18.54 & $9.54^{+0.15}_{-0.15}$ & $9.57^{+0.14}_{-0.15}$ & - & 272 \\ 
            1524 Joensuu & 15.52 & $15.51\pm0.009$ & $15.33\pm0.01$ & 0.18 & 15.39 & $9.25^{+0.17}_{-0.16}$ & $18.53^{+0.25}_{-0.27}$ & 9.276 & 260 \\ 
            2001 QO116 & 18.8 & $18.63\pm0.12$ & $18.54\pm0.16$ & 0.09 & 18.56 & $2.48^{+0.18}_{-0.16}$ & $7.43^{+0.11}_{-0.06}$ & - & 124 \\ 
            385 Ilmatar & 12.28 & $12.41\pm0.002$ & $11.90\pm0.002$ & 0.51 & 12.10 & $24.72^{+0.98}_{-1.75}$ & $27.48^{+1.46}_{-0.75}$ & 62.35 & 112 \\ 
            2014 KL4 & 18.91 & $19.04\pm0.16$ & $18.77\pm0.19$ & 0.27 & 18.87 & $8.61^{+0.19}_{-0.23}$ & $17.29^{+0.21}_{-0.55}$ & - & 111 \\ 
            1113 T-3 & 19.45 & $19.25\pm0.43$ & - & - & - & $5.77^{+0.07}_{-0.07}$ & $13.27^{+0.22}_{-0.15}$ & - & 91 \\ 
            2002 AC131 & 19.42 & $19.29\pm0.19$ & $19.08\pm0.34$ & 0.21 & 19.15 & $6.91^{+0.12}_{-0.06}$ & $20.80^{+0.92}_{-0.26}$ & - & 91 \\ 
            1998 FS53 & 19.35 & $19.35\pm0.42$ & $19.29\pm0.41$ & 0.06 & 19.30 & $1.98^{+0.09}_{-0.09}$ & $7.60^{+0.03}_{-0.07}$ & - & 77 \\ 
            19197 Akasaki & 18.26 & $18.37\pm0.08$ & $18.19\pm0.11$ & 0.18 & 18.25 & $5.81^{+0.75}_{-0.60}$ & $5.98^{+0.02}_{-0.04}$ & - & 71 \\ 
            1998 FN78 & 17.34 & $17.73\pm0.11$ & $17.34\pm0.05$ & 0.39 & 17.49 & $12.18^{+1.62}_{-1.28}$ & $10.88^{+0.22}_{-0.17}$ & - & 58 \\ 
            1999 CF55 & 18.60 & $18.42\pm0.09$ & - & - & - & $16.56^{+1.07}_{-1.80}$ & $7.26^{+0.03}_{-0.05}$ & 9.309 & 40 \\ 
            2002 CZ133 & 18.76 & $18.97\pm0.22$ & $18.58\pm0.27$ & 0.39 & 18.73 & $52.27^{+1.47}_{-1.39}$ & $22.99^{+0.43}_{-0.25}$ & - & 40 \\ 
            2001 VK58 & 19.41 & $19.08\pm0.65$ & $19.05\pm0.44$ & 0.03 & 19.04 & $23.25^{+0.18}_{-0.16}$ & $27.18^{+1.46}_{-0.86}$ & - & 36 \\ 
            2004 AG7 & 19.37 & $19.51\pm0.26$ & $19.20\pm0.35$ & 0.31 & 19.32 & $3.29^{+0.13}_{-0.12}$ & $6.66^{+0.06}_{-0.08}$ & - & 34 \\ 
        \bottomrule
        \end{tabular}
        \begin{tablenotes} 
            \item[a] These are the predicted magnitude in the \textit{V}-band from Lowell AstEph(\citealt{2022A&C....4100661M}).
            \item[b] $\textit{g}$ and $\textit{r}$ are obtained by averaging all photometric points in the \textit{g}-band and \textit{r}-band respectively. \textit{g - r} is the average color of the asteroids.
            \item[c] The $\textit{V}_\text{MST}$ are the results obtained by converting the observed magnitudes to the V-band using the formula $V = r + 0.44 \times (g - r) - 0.02$ (\citealt{1996AJ....111.1748F, 2002AJ....124.1776J})
            \item[d] The $\textit{P}_\text{LS}$ is the asteroid period obtained using the Lomb-Scargle method, while $\text{P}_\text{PDM}$ is obtained using the “phase dispersion minimization” method. The periods with more than 200 photometric points are considered reliable.
            \item[e] The periods data are from ALCDEF (\citealt{2011MPBu...38..172W}).
        \end{tablenotes}
        \end{threeparttable}
    \end{table}

\section{Discussion}
\label{sect:Discussion}

\subsection{Comparison with Existing Databases}

After the photometric analysis of the periods for the 22 asteroids in both the \textit{g}-band and \textit{r}-band, we obtain their magnitude and period. By averaging over all the photometric points of each asteroid, we get their average magnitude. We compared the average magnitude with the average magnitude predicted in the Lowell database (\citealt{2022A&C....4100661M}) which is used in Sec~\ref{sect:ephemeris} and the period with the ALCDEF\footnote{https://alcdef.org} 
(\citealt{2011MPBu...38..172W}), as shown in Table~\ref{Tab2}. ALCDEF is hosted by the NASA Planetary Data System (PDS), it stores raw asteroid time-series photometry and includes more than 24,000 asteroids.

It shows in Figure~\ref{Fig7}(b) that the asteroids with more than 200 photometric points obtained period data are very close to those in the ALCDEF. For the 22 asteroids analyzed in this work, periodic data from ALCDEF are available for 8 asteroids. Among them, 6 have more than 200 photometric points in this study, while the other 2 have inadequate data, making their periodic values unreliable. For the 6 asteroids with adequate photometric points, 5 show consistent 1:1 period matches with ALCDEF, while the other shows a 2:1 period. This discrepancy is likely due to the double-period processing applied in this study.

In our study, we converted the measured asteroid magnitudes to the V-band and compared them with the predicted magnitudes from Lowell Observatory. The average magnitude difference is only 0.14 mag, indicating good agreement with the Lowell database (see Table~\ref{Tab2}). Besides, MST has the advantage of multi-band photometry, we can get the color of the asteroids. After the subsequent multi-band photometry of more samples, the relationship between color and asteroid families (\citealt{1978AJ.....83..643D, 2002AJ....124.2943I})
can be analyzed more clearly and allow us to learn more about the composition of asteroids.

We also extract historical observations of 2000 NQ23 from ALCDEF including
\citet{2015AJ....150...75W} and \citet{2018MPBu...45..203R}.
Since the magnitude of an asteroid changes with distance, we use a simplified formula (\citealt{1978AJ.....83.1660G, 1989aste.conf..524B, 2004PhDT}) 
based on the relationship between the reduced magnitude $(\textit{M})$ \footnote{The reduced magnitude of the asteroid refers to the magnitude when the asteroid is 1AU away from both the Sun and the Earth, where the phase Angle$(\alpha)$ is not taken into account.} and observed magnitude $(\textit{M}_\text{obs})$ to convert magnitude for better comparison:

    \begin{equation}\label{eq2}
      \textit{M} = \textit{M}_{\text{obs}} - 5 \lg(\textit{R} \textit{$\Delta$}) \
    \end{equation}

where \textit{R} is the heliocentric distance, and \textit{$\Delta$} is the geocentric distance at the time of observation. Table~\ref{Tab3} shows the orbit parameters during the three observations of 2000 NQ23. The phase angle $(\alpha)$ correction is further performed according to \citet{1989aste.conf..524B} (see their equation (36)) to convert \textit{M} into absolute magnitude (\textit{H}). We apply the period analysis method employed in this work to the historical observational data, and the resulting periods are consistent with that reported ones in \citet{2015AJ....150...75W} and \citet{2018MPBu...45..203R} (see Table~\ref{Tab3}).

    \begin{table}[!ht]
        \centering
                \caption[]{ The observation of 2000 NQ23 }\label{Tab3}
        \begin{threeparttable} 
        \begin{tabular}{*{6}{c}} 
        \toprule
        \textbf{Telescope} & \textbf{Time} &\textbf{\textit{R}\text{(AU)}}\textsuperscript{a}& \textbf{\textit{$\Delta$}\text{(AU)}}\textsuperscript{a} & 
        $\textbf{\textit{P}}_\textbf{LS}\textbf{[h]}$\textsuperscript{b} & 
        $\textbf{\textit{P}}_\textbf{Ref.}\textbf{[h]}$\textsuperscript{c} \\
        \midrule
        1.22-m Palomar Schmidt\textsuperscript{d} & 09/10/2013 - 11/06/2013 & 2.054 - 2.219 & 1.207 - 1.352 & $4.228^{+0.001}_{-0.001}$ & $4.227\pm0.002$ \\
        0.35-m RMS Observatory\textsuperscript{d} & 08/24/2017 - 08/26/2017 & 1.872 - 1.873 & 0.881 - 0.888 & $4.22^{+0.03}_{-0.03}$ & $4.228\pm0.002$ \\
        MST & 12/28/2022 - 01/02/2023 & 2.912 - 2.947 & 2.003 - 2.126 & $4.23^{+0.03}_{-0.04}$ & - \\
        \bottomrule
        \end{tabular}
        \begin{tablenotes}
            \item[a] \textit{R} is the heliocentric distance, and \textit{$\Delta$} is the geocentric distance at the time of observation.
            \item[b] $\textit{P}_\text{LS}$ is the period derived from historical data using the Lomb-Scargle method employed in this work.
            \item[c] $\textit{P}_\text{Ref.}$ is the period reported in the literature (\citealt{2015AJ....150...75W, 2018MPBu...45..203R}).
            \item[d] The observation data are from ALCDEF (\citealt{2011MPBu...38..172W}).
        \end{tablenotes}
        \end{threeparttable}
    \end{table}

\subsection{Databases of Asteroid Light Curves}

There are some important databases of asteroid light curves and rotation periods, such as the Asteroid Lightcurve Database (LCDB), and ALCDEF, etc. Similar databases and studies exist for specific missions such as the Transiting Exoplanet Survey Satellite (TESS) and the Wide-Field Infrared Survey Explorer (WISE). LCDB is the most authoritative repository of asteroid lightcurve parameters and other information, such as estimated/measured diameters, absolute magnitudes \textit{(H)}, phase slope parameters \textit{(G)}, albedos, and more
(\citealt{2009Icar..202..134W}).
LCDB has data for 341,131 asteroids, but only 36,259 of them have periodic analyses with at least two different observations\footnote{https://minplanobs.org/mpinfo/php/lcdb.php}. LCDB does not store the original data used to do that analysis but ALCDEF does.  \citet{2020ApJS..247...26P} analyzed and processed the TESS-DR1 data, obtained the light curves of 9912 asteroids.
\citet{2022PSJ.....3...30M} analyzed some of WISE's high-quality data, obtained the light curves of 4412 asteroids. However, TESS and WISE do not have a more comprehensive database of asteroid light curves that is updated in time. So the amount of available light curve data for asteroids remains significantly smaller compared to the total number of known asteroids, which exceeds 1,340,000 (see Sec.~\ref{sect:ephemeris}). These databases are listed in Table~\ref{Tab4}.

    \begin{table}[!ht]
        \centering
            \caption[]{ Lightcurve Database of Asteroids }\label{Tab4}
        \begin{threeparttable} 
        \begin{tabular}{*{5}{c}} 
        \hline
            \textbf{Database} & \textbf{Number} & \textbf{Datebase updated} & \textbf{Reference} \\ \hline
            LCDB & 36259 & 10/2023 & \citet{2009Icar..202..134W} \\ 
            ALCDEF & 24450 & 04/2024 & \citet{2011MPBu...38..172W} \\ 
            TESS DR1 & 9912 & 01/2020 & \citet{2020ApJS..247...26P} \\ 
            WISE & 4420 & 02/2022 & \citet{2022PSJ.....3...30M} \\
            CNEOST & 506 & 07/2020 & \citet{2020AJ....160...73Y} \\ \hline
        \end{tabular}
        \end{threeparttable}
    \end{table}

According to the results of this work, three key factors underscore the exceptional potential of the SiTian project in asteroid photometry. Firstly, the SiTian project will scan at least 10,000 square degrees of sky in half an hour (\citealt{2021AnABC..93..628L}). The FOV is much larger than that of MST, dramatically increasing observational coverage. Secondly, its detection limit of 21 magnitude (\citealt{2021AnABC..93..628L}) outperforms the 20 magnitude threshold established in this work, enabling the observation of fainter asteroids. Lastly, the ecliptic latitude of this work is $18.37^\circ$, observations at lower ecliptic latitudes promise greater asteroid densities, considering that 96\% of known asteroids reside within a $20^\circ$ band around the ecliptic plane (\citealt{1994IAUS..161..385M, 2024Master}). Integrating these advantages, preliminary estimates that the SiTian project can complete the photometry of one hundred thousand asteroids within one month of observation. A more detailed evaluation and simulation of the asteroid research capabilities of the SiTian project (considering various observation modes such as long-time monitoring, all-sky surveys, and ecliptic plane surveys) is in progress. 

\section{Conclusion}
\label{sect:conclusion}
In this work, a pipeline is prepared for asteroid photometry in the SiTian project. Using this pipeline we complete the photometry for 22 asteroids with $V \leq 20 \, \text{mag}$ in the MST f02 region and obtain light curves and periods. After comparison with ALCDEF, the period of asteroids with photometric points greater than 200 matches well, which verifies that MST can perform the photometric work of $V \leq 20 \, \text{mag}$ asteroids. We present the procedure of data processing in detail.

This work utilizes the SiTian pathfinder project, MST, to demonstrate its potential for advancing research on small Solar System bodies. MST’s wide-field observational capability enables the collection of extensive data for constructing light curves, determining rotation periods, and studying asteroid shapes and sizes in the future. These preliminary efforts aim to enhance databases like ALCDEF, deepen our understanding of asteroid properties, and support future spectroscopic and dynamical studies.

\begin{acknowledgements}
The SiTian project is a next-generation, large-scale time-domain survey designed to build an array of over 60 optical telescopes, primarily located at observatory sites in China. This array will enable single-exposure observations of the entire northern hemisphere night sky with a cadence of only 30-minute, capturing true color (gri) time-series data down to about 21 mag. This project is proposed and led by the National Astronomical Observatories, Chinese Academy of Sciences (NAOC). As the pathfinder for the SiTian project, the Mini-SiTian project utilizes an array of three 30 cm telescopes to simulate a single node of the full SiTian array. The Mini-SiTian has begun its survey since November 2022. The SiTian and Mini-SiTian have been supported from the Strategic Pioneer Program of the Astronomy Large-Scale Scientific Facility, Chinese Academy of Sciences and the Science and Education Integration Funding of University of Chinese Academy of Sciences.

We sincerely thank the reviewer for the constructive comments and suggestions, which have significantly improved the quality of this manuscript. We are grateful to Kai Xiao, Jing Wen, and Shuai Feng for very helpful discussions. J.G acknowledgments the supports from the National Natural Science Foundation of China (NSFC) under Nos. 12203002 and 11973015; Y.H acknowledgments the supports from the National Natural Science Foundation of China (NSFC) under 12422303 and 12261141690, the supports from the National Key Basic R\&D Program of China via 2023YFA1608303 and the Strategic Priority Research Program of the Chinese Academy of Sciences (XDB0550103); S.M.H acknowledges the supports from National Natural Science Foundation of China (NSFC; grant Nos. 12373015); H.Z acknowledges the supports from National Natural Science Foundation of China (NSFC; grant Nos. 12120101003 and 12373010) and National Key R\&D Program of China (grant Nos. 2023YFA1607800, 2023YFA1607804, 2022YFA1602902), Beijing Municipal Natural Science Foundation (grant No. 1222028), Strategic Priority Research Program of the Chinese Academy of Science (Grant Nos. XDB0550100 and XDB0550000).
\end{acknowledgements}

\bibliographystyle{raa}            
\bibliography{RAA-2024-0356.R3}

\end{document}